# System Tests of the ATLAS Pixel Detector

K. Reeves[1], K.-H. Becks[1], T. Flick[1], P. Gerlach[1], J. Grosse-Knetter[2], F. Hügging[2], M. Imhäuser[1],
S. Kersten[1], P. Kind[1], K. Lantzsch[1], P. Mättig[1], I. Rottländer[2], J. Richter[1], J. Schultes[1], J. Schumacher[2],
J. Weingarten[2], and N. Wermes[2]
on behalf of the ATLAS Pixel Group

[1]Bergische Universität Wuppertal, Fachbereich C, Gaußstr. 20, D42097 Wuppertal, Germany
kreeves@physik.uni-wuppertal.de

[2]Universität Bonn, Nussallee 12, D53115 Bonn, Germany

*Abstract*

The innermost part of the ATLAS (A Toroidal LHC ApparatuS)[1] experiment at the LHC (Large Hadron Collider) will be a pixel detector, which is presently under construction. Once installed into the experimental area, access will be extremely limited. To ensure that the integrated detector assembly operates as expected, a fraction of the detector which includes the power supplies and monitoring system, the optical readout, and the pixel modules themselves, has been assembled and operated in a laboratory setting for what we refer to as system tests. Results from these tests are presented.

## I. INTRODUCTION

The Pixel Detector is the component of the ATLAS experiment[1,2] located closest to the interaction region. It is comprised of a barrel section and two identical disk sections located on opposing sides of the primary interaction point (see **Figure 1**). Each of these sections has three layers, ensuring that the pixel detector will provide three space-points per track for nearly all tracks with a pseudorapidity magnitude less than 2.5.

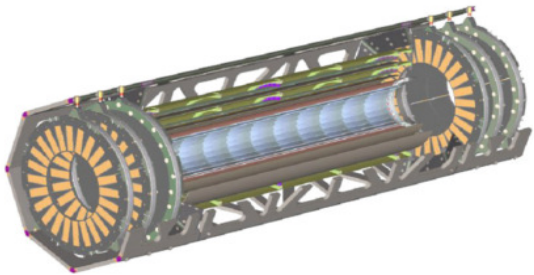

Figure 1 Schematic drawing of the ATLAS Pixel Detector, depicting the three layers of the barrel section as well as the two disk sections.

### A. Modules

The most basic element of the pixel detector is the pixel module. Each module is comprised of a sensor of oxygenated silicon which is bump-bonded to 16 front end chips for readout. These front end chips are wire-bonded to a flex circuit which has been glued to the opposite side of the sensor. The flex circuit provides the mounting point for the Module Controller Chip (MCC) [3], which provides the communication to the front end chips [4] on the module, as well as module-level event building. In addition, for the barrel modules, the flex circuit provides the connector to which a cable will be connected, providing communication and powering pathways (see **Figure 2**). Each module has 46,080 pixels measuring 50 x 400 μm[1], and there will be 1744 modules in the detector.

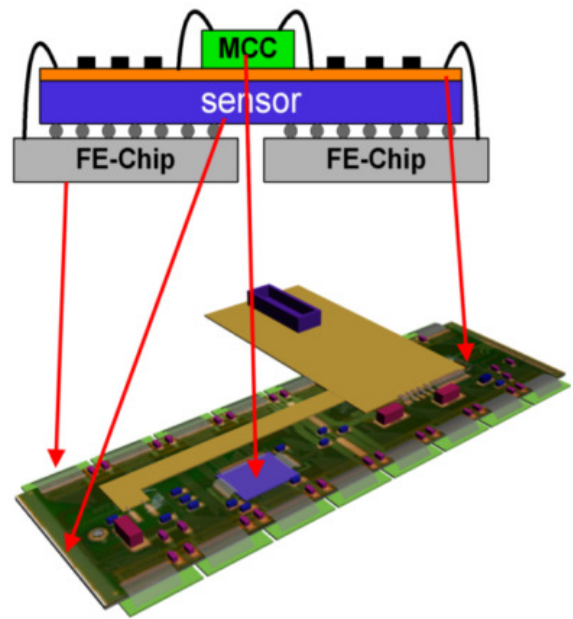

Figure 2 Diagram relating a cross-sectional view to the physical layout of a pixel module, which measures roughly 2x8 cm.

As the front end chips have both analogue and digital circuits requiring two different supply voltages, those must be provided separately for powering and in addition to the high voltage used for sensor depletion.

---
[1] A small fraction of pixels located along the edges of the front end chips have different geometries to avoid insensitive areas.

### B. Disks

For the disk sections, the basic mechanical unit is referred to as a sector (see **Figure 3**). Each sector has mounted upon it 6 modules, 3 on each side and arranged to ensure that the unavoidable gaps in coverage on a given side are covered by the modules on the opposing side. The sectors are constructed from carbon composites, with cooling provided by means of a tube interleaved throughout the sector volume. Each disk will be constructed from 8 sectors, and so shall carry 48 modules. All disk modules are read out at 80 Mbit/s data transfer rates.

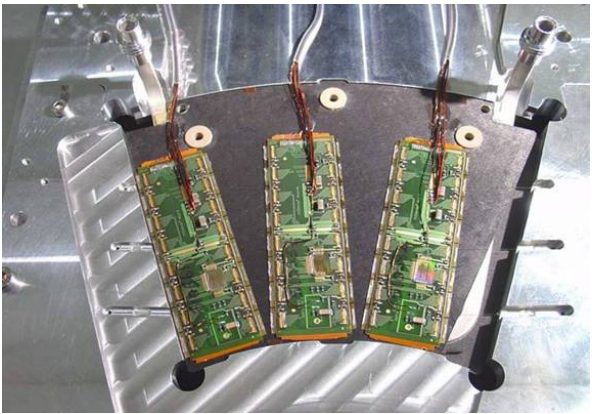

Figure 3 A sector with 3 modules visible. On the opposite side, 3 additional modules are arranged to cover the gaps between those seen here.

### C. Barrel Section

The basic mechanical unit used in the barrel section is referred to as a stave (see **Figure 4**). Each stave is formed from a carbon composite structure which will have 13 modules glued onto its surface. Cooling of the modules is provided by means of a thin-walled aluminum tube attached to the underside of the stave.

The three barrel layers are, moving from smaller radii to larger radii, the B-layer, layer 1, and layer 2. The innermost B-layer, at a radius of 50.5 mm, is comprised of 22 staves carrying 286 modules. The proximity of this layer to the interaction region results in the highest occupancies for the detector; as a result, the B-layer is read out at 160 Mbit/s. Layer 1, at 88.5 mm, is an assembly of 38 staves, for 494 modules. It will be read out at 80 Mbit/s. The outermost layer 2, at 122.5 mm, has 52 staves providing 676 modules, and will be read out at 40 Mbit/s.

### D. Readout

Each module is connected by means of a fine-wire aluminum cable (type-0, see **Figure 4**) to a patch panel (PP0) located approximately one meter from the interaction point. An optical-electrical conversion board, which is referred to as an optoboard, is mounted on the PP0 and there provides conversion of outgoing electrical signals to an outgoing optical signal, as well as converting incoming optical transmissions to electrical signals which are forwarded to the modules over the type-0 cable. The optoboard is connected via approximately 80 meters of optical fiber to matched opto-electric components [5] which are mounted on a Back-of-Crate (BOC) card which is mounted in a pixel readout crate in the ATLAS electronics trailer. The BOC also provides the fiber-optic connection to the ATLAS data acquisition system by means of an onboard fiber optic receiver card. The BOC is connected to a Readout Driver (ROD) [6] through the backplane of the crate. The ROD provides the computational capabilities necessary for handling the vast amount of data which must be processed to calibrate the Pixel Detector, in addition to event building, error handling, and data formatting.

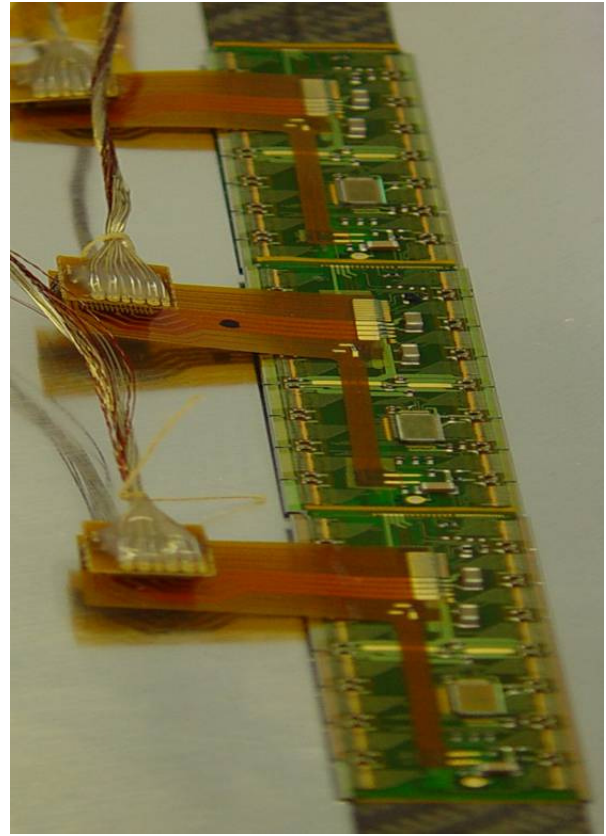

Figure 4 Photograph of a stave prototype. Three modules can be seen to have been glued onto a carbon composite structure, and cables – referred to as type-0 cables – providing powering and data transmission pathways have been connected.

### E. Detector Control System (DCS)

The DCS – as currently used – is responsible for powering the modules and the on-detector optical components, monitoring critical voltages, currents, and temperatures, and responding to conditions where values are outside of the acceptable ranges. In addition, graphical interfaces providing the user with an overview of detector conditions and the ability to effect desired changes in the state of the detector are provided.

## II. THE SYSTEM TEST SETUP

All components of the detector are thoroughly tested and qualified prior to being approved for final usage in the detector. However, the testing regimen – by necessity – employs equipment which is substantially different than what shall finally be used, leaving open questions of interoperability of production parts. The system test has been designed to verify that the pixel modules perform as during the production testing when connected to the production powering and readout components, and operated in a parallel fashion.

### A. Pixel Modules

Several staves have been produced while the sites which will be performing the module mounting have verified the methods to be employed. Some of the earlier of these staves were equipped with preproduction modules which have been constructed with FE-I2.1 readout chips, which are one revision prior to the production versions, FE-I3. It should be noted that the differences between these two revisions are rather minor, and should have no impact on the studies discussed here.

Two of these staves with preproduction modules have been assembled into a bi-stave[2] and dedicated to system testing, providing 26 modules for study [7].

### B. Detector Control System (DCS)

An integral part of the system test has been to build up a first complete version of the pixel detector DCS employing current versions of the control and monitoring software, in addition to the various hardware elements listed below:

- Wiener power supply providing the module low voltages (commercial)
- iseg High Voltage Power Supply (commercial)
- SC-OLink providing on-detector optical component powering (developed at University of Wuppertal)
- Module voltage regulator boards (PP2, developed at INFN, Milan)
- Production length power cables for ½ of one stave
- Thermal regulation

For a more complete description of the DCS system, see [8].

---

[2] A bi-stave is comprised of two staves which have been joined mechanically, and which share a common cooling circuit. All staves will be joined into bi-staves prior to being mounted into the final detector assembly.

### C. Communication and Readout

The interface to the detector assembly is provided by means of a VP-110 Single Board Computer (SBC), a commercial crate controller from Concurrent Technologies. Communications between the SBC and the ROD is over the crate VME bus, with the ROD forwarding commands to modules through the interconnected BOC [9]. Data returning from the modules is then passed through the BOC to the ROD, where it can either be processed with resultant data to be forwarded to the SBC, usually in the form of histograms, or the raw data can be forwarded for off-board analysis.

For system testing a Revision E ROD has been used, which is one revision prior to the production Revision F. As the only difference between the two revisions is the VME interface, this is expected to have no impact on these studies. While the BOC being used is a preproduction version, there have been no substantive design changes for the production version, and so this also should have no impact on these studies.

## III. DATA AND ANALYSIS

The primary method employed in these tests was a threshold scan, referred to here as a VCAL scan (see Figure 5). By fitting an error function to the scan data, threshold and noise values can be obtained and then compared to values previously obtained.

An example of threshold values obtained by such a VCAL scan can be seen in **Figure 6**, which provides three different views of the threshold distributions determined for the module. The upper plot shows a topographical distribution of the pixels, with the color (or shading) representing the threshold value, which is useful for observing localized effects, such as masked pixels which appear as white areas. The two histograms are self-explanatory.

**Figure 7** shows the noise distributions obtained during the VCAL scan. As with the threshold scans just discussed, the upper plot provides a topographical representation of the noise values obtained. The one-dimensional histograms show the results for all pixels on the module (*Noise distribution*) and for the various pixel geometries needed to avoid insensitive areas at the edges of the readout chips (*Noise distribution long*, *Noise distribution ganged*, and *Noise distribution inter-ganged*). The module is seen to be well-tuned from the uniformity of the threshold and noise values.

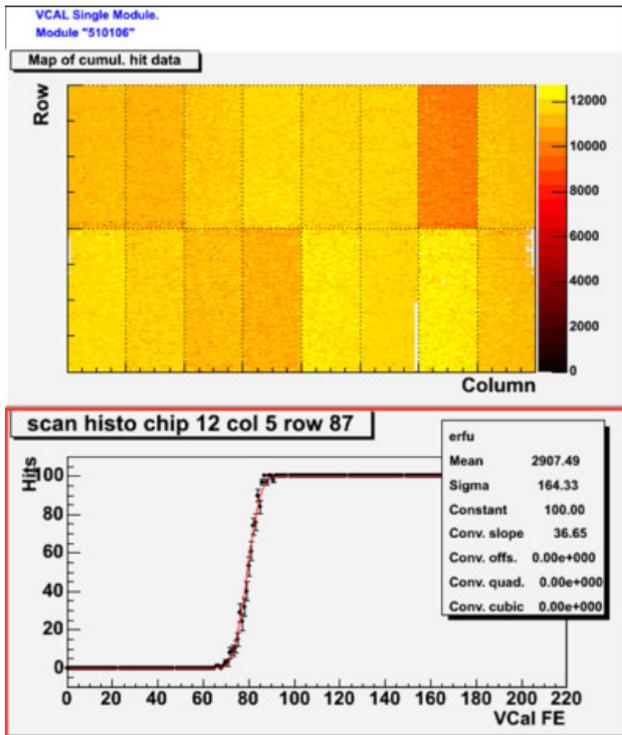

Figure 5  Example of the output of a VCAL scan for module 510106. The 16 readout chips can be clearly distinguished in the upper plot.

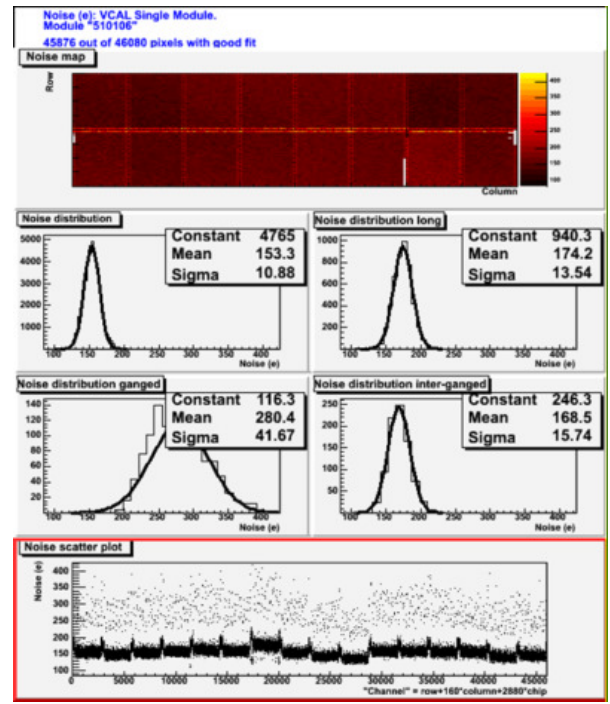

Figure 7  Noise distributions. As with Figure 6, the upper plot provides a topographical view of the noise for individual pixels.

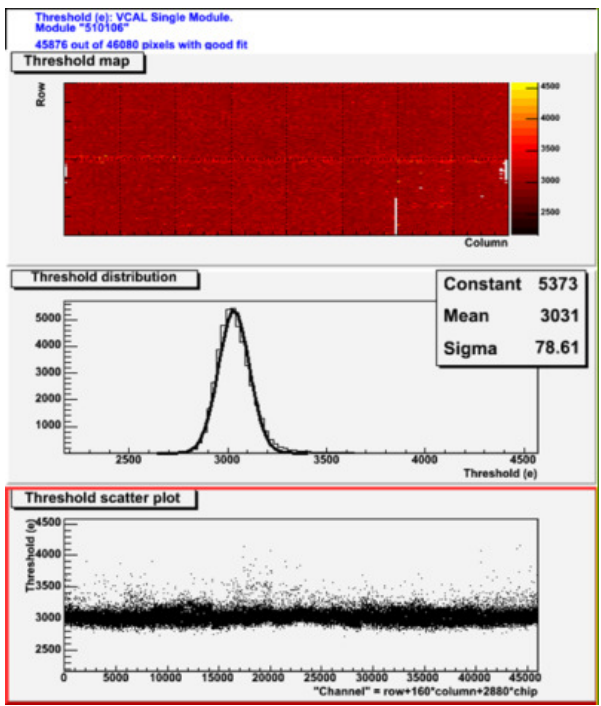

Figure 6  Threshold distributions. The upper plot provides a topographical view of the individual pixel thresholds, (the white areas indicate masked pixels).

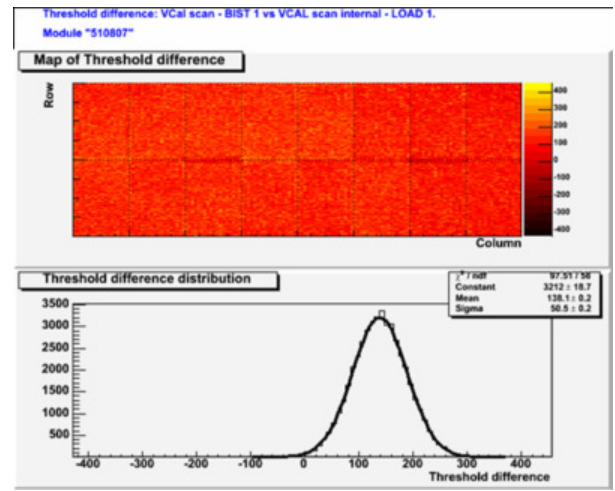

Figure 8  Threshold difference plot obtained from threshold values obtained during production testing subtracted from a system test scan. The offset in the mean value of 138.1 is attributed to different operational temperature.

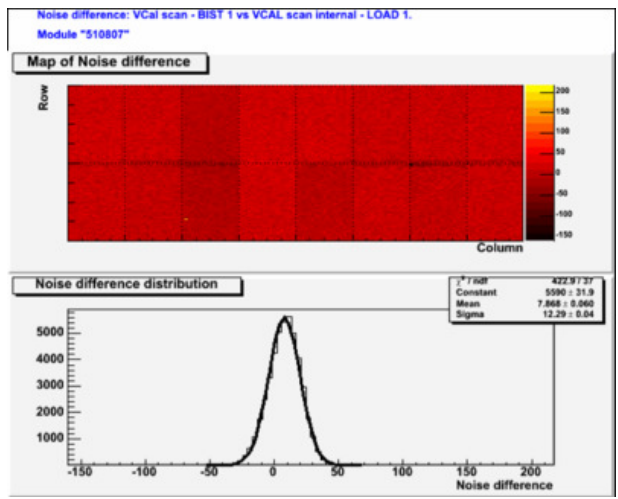

Figure 9  Noise difference plot, from the same data sets as with Figure 8.

One can study the consistency of data obtained in two different VCAL scans by performing a difference comparison of the resultant data. An example of a threshold difference analysis can be seen in **Figure 8**. One sees from these plots that the threshold values are quite consistent between the two scans, although there is an offset of 138.1 electrons. This is attributed to the different temperatures at which the two scans were obtained, with the first scan being conducted at a temperature near 19C, while the second scan was obtained at a temperature near 25C. Further tests will be conducted in order to eliminate this thermal effect.

The difference analysis can be applied to the noise values, as in **Figure 9**, which shows clearly that no major effects have been introduced by the transition from electrical readout to optical readout, or from the use of testing power supplies to design power supplies [7,8]. As with the threshold analysis, the shift in the mean of the distribution is attributed to obtaining the data at different operational temperatures.

## IV. CONCLUSIONS

The system test studies thus far completed confirm the basic interoperability of the many components used for powering and readout of the ATLAS pixel detector. The data thus far obtained shows no effect from powering the modules with the design power supplies, nor from reading them out over the optical data path. These studies validate the design for 40 Mb/s operation, and tests expected to be completed in the near future are expected to validate the 80 Mb/s and the 160 Mb/s modes. In addition, a large system test is being assembled at CERN which will present an opportunity to repeat these tests with approximately 10% of the modules to be deployed in the experiment.

## V. ACKNOWLEDGEMENTS

The experiences described here have only been possible due to the work of all members of the ATLAS pixel group. It is a pleasure to have the opportunity to offer our thanks to each of them for these efforts, and for contributing to such a pleasant collaboration.

We gratefully acknowledge the funding support of the Bundesministerium für Bildung und Forschung.